# Decision Making by a Neuromorphic Network of Volatile Resistive Switching Memories

Saverio Ricci*, David Kappel†, Christian Tetzlaff†, Daniele Ielmini*, Erika Covi‡
*Dipartimento di Elettronica, Informazione e Bioingegneria, Politecnico di Milano, Milano, Italy.
†University of Göttingen, Göttingen, Germany.
‡NaMLab gGmbH, Dresden, Germany
corresponding author: saverio.ricci@polimi.it

*Abstract*— The necessity of having an electronic device working in relevant biological time scales with a small footprint boosted the research of a new class of emerging memories. Ag-based volatile resistive switching memories (RRAMs) feature a spontaneous change of device conductance with a similarity to biological mechanisms. They rely on the formation and self-disruption of a metallic conductive filament through an oxide layer, with a retention time ranging from a few milliseconds to several seconds, greatly tunable according to the maximum current which is flowing through the device. Here we prove a neuromorphic system based on volatile-RRAMs able to mimic the principles of biological decision-making behavior and tackle the Two-Alternative Forced Choice problem, where a subject is asked to make a choice between two possible alternatives not relying on a precise knowledge of the problem, rather on noisy perceptions.

## I. Introduction

Experimental psychology and neuroscience have made great progress in understanding the mechanisms that equip the brain to make decisions under uncertainty [1]. Porting these decision-making strategies to automated and Artificial Intelligence (AI) systems has gained strong interest in the last few decades [2]. Algorithmic decision-making is a special field of AI that strives to develop intelligent systems able to process the input data to produce a score or a choice that is used to support decisions [3] such as prioritization, classification, association, and filtering. In this context, the ability of an artificial system to work at biologically relevant timescales, i.e., in the order of milliseconds to seconds, is of utmost importance [4]. However, in current standard CMOS technology, these time constants are obtained using rather area-consuming capacitors that are charged and discharged. Recently, silver-based RRAMs [5-9] have been proved to feature similarity to the biological synapses characterized by a short-term dynamic. Together with their nanoscale footprint and high ON/OFF ratio, these devices are really appealing in AI systems. Indeed, the information is stored in the physical configuration of the device, i.e., a conductive filament inside the oxide. Therefore, the short-term state of the network is stored as a hidden variable directly in the RRAM, without need for charging or discharging currents.

In this work, we present volatile resistive switching memories based on hafnium oxide and silver as a suitable elementary unit for brain-inspired architectures and neuromorphic tasks and we use them to implement a simple two-alternative forced choice (2AFC) algorithm, where a binary decision has to be made based on noisy observations. First, we show the main electrical properties of the single RRAMs, in terms of switching voltage thresholds and retention capabilities, which are two key parameters in a system able to implement 2AFC tasks. Multiple devices are then studied to understand the behavior of a population and its electrical properties. Finally, a 2AFC task is presented and several simulations are carried out to investigate the behavior of the neuromorphic system.

## II. Device Characterization

### A. Electrical properties of volatile RRAMs

RRAMs are two-terminal resistive elements with memory effects, where the resistance of the device depends on the internal distribution of metallic cations that migrate from the silver electrode into the oxide and for a conductive filament (CF) [6]. They belong to class of filamentary switching memories [6, 7,10]. Fig. 1a sketches the internal structure of the device: when a CF is formed, the resistivity of the device is in the order of magnitude of the k$\Omega$. On the other hand, if a gap exists between the electrodes the resistance is high, in the order of the T$\Omega$. The silver metal creates a CF made of Ag$^+$ cations inside the oxide and it has been proved to create a volatile behavior [5-10], i.e., the conductive filament is spontaneously dissolved in absence of an electric field. Here we propose a 1-resistor 1-transistor structure, where the 1R is the RRAM and the 1T is a foundry-level FET used as a current limiter to better tune the device properties [6,8]. Fig. 1b reports a quasi-static IV sweep curve, where a sudden transition from low to high current happens when the voltage overcomes a threshold, called set voltage, creating the CF. When the voltage decreases, the filament spontaneously disrupts and the current drops back to low values. The dissolution time depends on the diameter of the filament, which depends on the current flowing through the device [10]. The same happens in the negative region. To better understand the temporal evolution of the volatile memristors, a semi-triangular pulse is applied to switch the devices ON and then the status is monitored with a -150 mV voltage to avoid any influence of the read voltage on the device, as shown in Fig. 1c, revealing the RRAM spontaneously switches OFF after a time called retention. During the semi-triangular pulse, the device switches on and the current reaches the compliance, fixed with the 1T in the saturation region, applied both to protect the device from permanent short-circuit and to control the RRAM properties. Fig. 1d shows the distributions of the set voltage for different compliance currents ($I_{CC}$). As expected, there is no dependence of the set voltage on the $I_{CC}$ since the CF formation has no memory of the previous operations [5-8]. The distributions also suggest the possibility of using the voltage amplitude of the programming pulse to tune the switching probability ($P_{ON}$). The retention time, in Fig. 1e, shows a dependence on the $I_{CC}$. The average retention time can be modulated by the current compliance [6,8,10]. Fig. 1f highlights the switching probability as a function of the pulse amplitude.

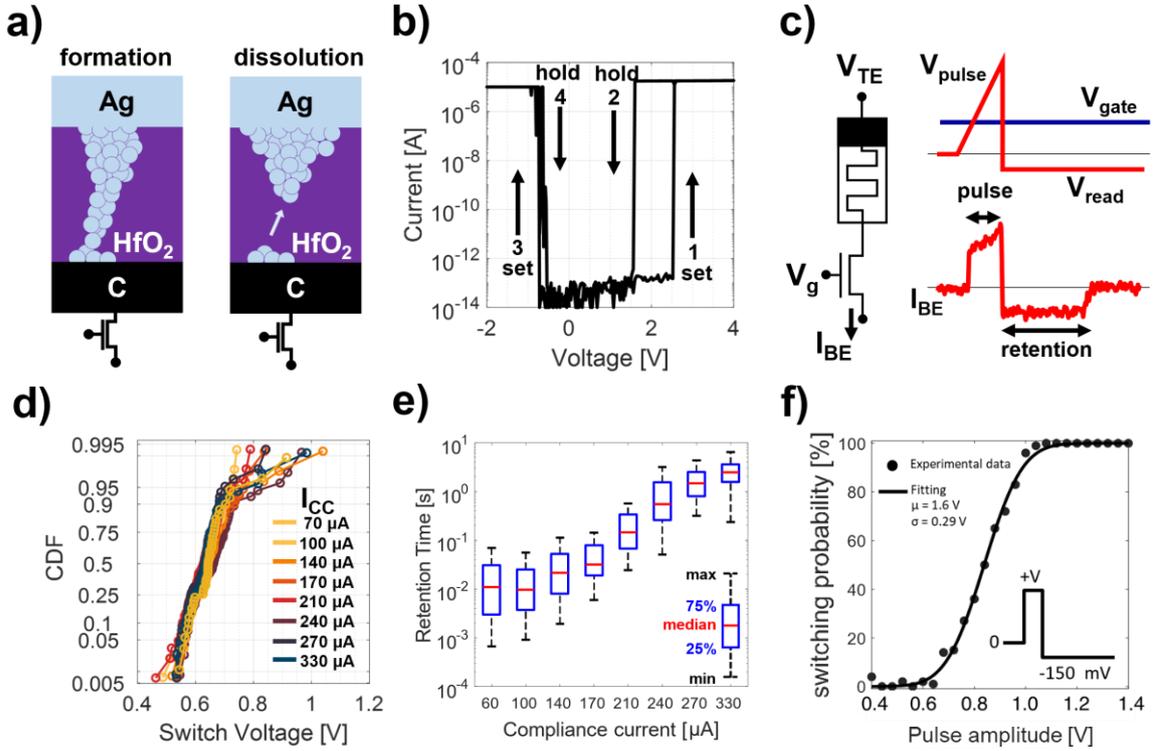

Fig. 1: a) Sketch of the device with the 1T1R structure and working principle with the filament formation (set) and filament spontaneous dissolution (reset). b) Quasi-statical voltage-current curve with 10 µA of compliance current ($I_{CC}$). c) Temporal characterization to study the spontaneous dissolution of the conductive filament. A semi-triangular pulse is applied to switch the device ON and a subsequent read voltage of -150 mV is applied to monitor the device behavior. d) Distribution of the set voltage for different $I_{CC}$. e) Box plot of the retention time as a function of the compliance current. The median value is in red. The box collects the values between 25% and 75%. f) Switching probability as a function of $V_{set}$.

*B. Synapse with multiple devices*

In this section, the behavior of a synapse composed by multiple devices in parallel is presented, as a way to overcome the binary dynamics of a single RRAM in favor of a more gradual behavior [8,9]. Fig. 2a represents such architecture with parallel devices, with all the 1T1R structures sharing the correspondent electrodes. When a pulse is sent, the RRAMs switch ON depending on their switching probability and the overall synaptic current is proportional to the number of ON devices. At the end of the stimulation, the devices switch OFF depending on their retention time, resulting in an exponential behavior [9,10]. Fig. 2b analyses the impact of the $P_{ON}$ on a 50-RRAM based system after the application of 50 spikes at 10 Hz (sketched as black bars above the curves) in which the $I_{CC}$ is 300 µA. The system shows a linear integration behavior for small probabilities ($P_{ON} < 5\%$) and a saturation region for $P_{ON} > 10\%$, where all the devices are ON. The temporal dynamics can be tuned by changing the $I_{CC}$. Fig. 2c, for example, compares the temporal response of the system both to the stimulation (10 Hz, $P_{ON} = 10\%$) and to the relaxation. Since a small $I_{CC}$ leads to a short retention time, the single devices switch OFF quickly and the overall current stays low. With an increase of $I_{CC}$, the average current increases with the number of pulses. For a proper system operation, therefore, it is important to select a $P_{ON}$ that do not lead to an early switch ON of all devices, as well as a $I_{CC}$ that allows for retention times matching the relevant timescale of the task at hand. The multi-device synapse allows both features with an additional gradual change of the current that is beneficial in neuromorphic systems [9].

## III. TWO-ALTERNATIVE FORCED CHOICE

Two-alternative forced choice (2AFC) is a special class of design-making algorithm which plays a crucial role in the biology [1-3]. The importance of the 2AFC finds its root in the necessity of choosing between two options without considering the exact evolution of an environment [3], but looking at the noisy observations gained in time, thus the choice is made with a final perception. Experimental studies suggest that humans and animals use internal neural representations for evidence integration to solve 2AFC (and related) tasks [1]. Decisions are based on internal representations for evidence integration rather than explicit counting. The concept of the 2AFC is therefore used as a benchmark to study the accuracy of a real, or an artificial, system that has to decide between two options. Fig. 3a explains the most basic 2AFC experiment and how it works. Two random input streams coming from two sources A and B (e.g., two flashing lights of different mean frequency) are sent. At the end of the streams, the system has to indicate which one of the two sources fired the most [2]. Fig. 3a illustrates the internal hidden representation that is used for evidence integration (orange traces). When a stimulus arrives the value of the variable increases while, if nothing arrives, it slowly decays. Comparing the two results, the choice is made.

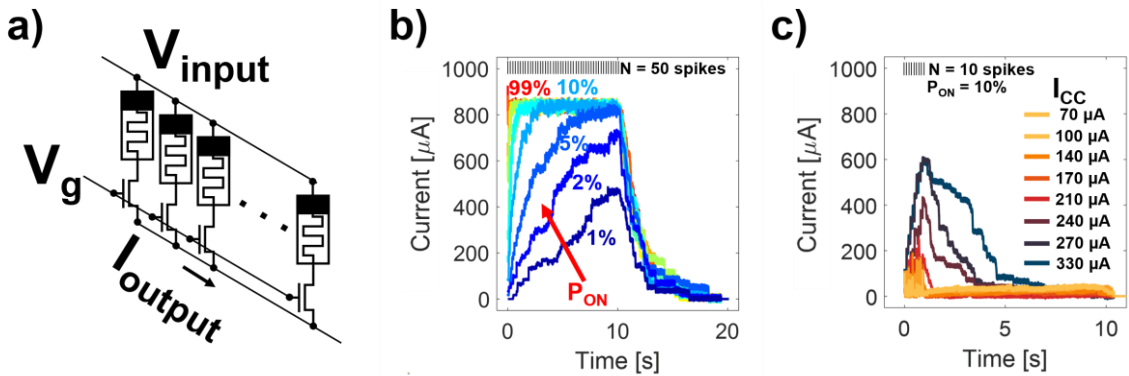

Fig. 2: a) Sketch of the multi-device synapse architecture, with parallel RRAMs devices. b) Impact of the switching probability ($P_{ON}$) on the synaptic current. Increasing the $P_{ON}$ the neuron response is faster and reaches the saturation, where all the devices are ON. c) Impact of the current compliance in the synaptic current.

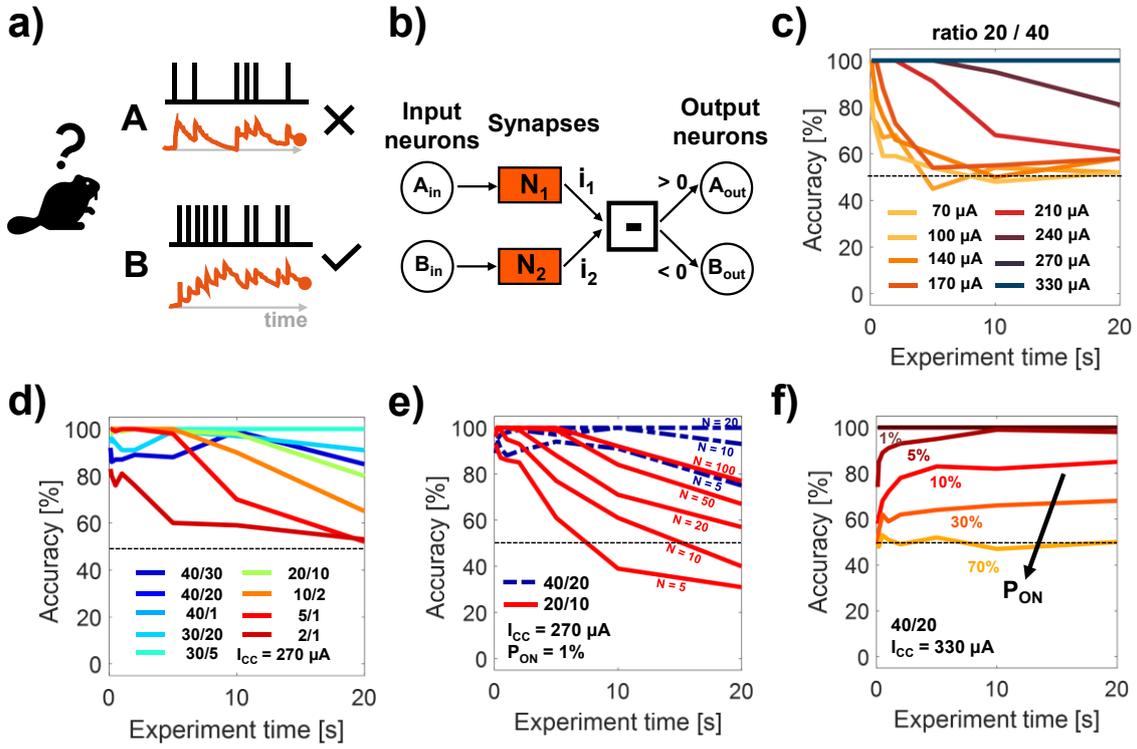

Fig. 3: a) Sketch of the Two-Alternative Forced Choice experiment. In the experiment, two streams containing a different number of stimuli are sent to the studied subject which is asked to decide between the two streams. b) Scheme of the proposed architecture: input neurons (A and B) collect the signals and fire the RRAM-based synapses. The synaptic currents are then collected by the corresponded output neurons and the neuron with the highest current is activated. c) Accuracy of the system depending on the $I_{CC}$. d) Effect of different spike rates on the system accuracy. e) Comparison of the accuracy at different number of devices in the synapse and at different spike densities. f) Impact of the switching probability on the accuracy.

Here we propose a system, sketched in Fig. 3b, composed of an input neuron layer which fires to the synapses $N_1$ and $N_2$. The volatile synapses are implemented with a variable number of devices (N), characterized by a certain retention time distribution depending on the current compliance ($I_{CC}$) and a certain switching probability ($P_{ON}$) related to the spike voltage amplitude. The synapses integrate the signal coming from the input neurons. The difference of the two currents $i_1-i_2$ is sent to the output layer (neurons $A_{out}$ and $B_{out}$). The comparator can be, as an example, implemented using a winner-take-all circuit [11] to optimise the occupied area. If the resulting current is positive (negative), neuron $A_{out}$ ($B_{out}$) will fire, thus choosing the stream from $A_{in}$ ($B_{in}$) as the one who fired the most. The streams have a different number of pulses and they are randomly distributed during the simulation time. We define the accuracy as the number of correct choices with respect to the total number of attempts. To study different biological scenarios, multiple simulations are run using different stimulation times (ranging from 500 ms to 20 s), different number of spikes per stream, and different ratio between the number of spikes fired by the two input neurons. Fig. 3c compares the accuracy in the choice between two streams with 20 and 40 pulses (ratio 40/20) sent within different simulation times and changing the $I_{CC}$. When the average retention time of the volatile synapse is shorter than the simulation time (meaning the pulses are more spaced) the accuracy decreases. The accuracy at 50% means that the system is randomly choosing between the two streams. Therefore, we confirm the importance of matching the time constants of the system to the one of the task at hand.

Afterwards, we study the system using a fixed compliance current and varying the ratio between spikes in the streams and their number, as reported in Fig. 3d. Given the same ratio, a higher number of pulses lead to a higher accuracy (see 40/20, 20/10, and 2/1 data in Fig. 3d), since more experience can be collected and more data integrated. The same is true for streams with different ratio: the higher the ratio, the higher the difference between the currents $i_1$ and $i_2$, and the higher the accuracy. This behavior is in good agreement with the results obtained by Brunton et al. [1], where the subjects are more sensitive to streams with a large number of stimuli and a great mismatch between the two streams [2, 3]. Interestingly, in our system we also note that when the number of stimuli is limited, e.g., 2/1, 5/1, and 10/2 in Fig. 3d, the time of the simulation plays a crucial role. Indeed, the accuracy is higher when the duration of the experiment is short. This is an effect of the short-term dynamics of the devices, i.e., the system tends to forget if not stimulated for a long time.

To evaluate the impact of the number of devices constituting a synapse in our artificial system, the number of elements was changed. In Fig. 3e two different conditions are presented, using a number of elements from 3 to 100 and an $I_{CC} = 270$ µA. For the 40/20 ratio, 20 RRAMs are sufficient to ensure an almost always correct choice. With short time duration experiments also 5 devices are enough to ensure an accuracy greater than 90%. The situation changes when the ratio is 20/10. Despite for short time scales there is a small difference in the accuracies, when the experiment duration increases, the larger the number of elements and the higher the ability to choose correctly. In addition to the effect of the number of elements on the accuracy, the current compliance plays also a crucial role in the system. There is a critical number of elements, which depends on all the parameters (like $I_{CC}$ and $P_{ON}$), after which the system is saturated. Considering all the parameters explored in the simulations, a number of elements between 30 and 50 is enough to ensure an accuracy higher than 60% for all the simulation times. Finally, we also studied the dependence on $P_{ON}$ on the system, as shown in Fig. 3f. The increase of the switching probability (as for $P_{ON} = 5\%$) results in a faster activation of the devices, as already discussed in Fig. 2b. As a consequence, in the proposed system we evidence a decrease of the accuracy with an increase of $P_{ON}$. This result supports the importance of a probabilistic approach to improve the performance and robustness of the system accomplishing decision-making tasks. This result supports the importance of variability in the devices to improve the robustness of the system accomplishing decision-making tasks.

## IV. CONCLUSION

Volatile RRAMs based on silver and hafnium oxide feature spontaneous switching dynamics with biologically relevant time constants. The devices are characterized by two different resistive states, a switching probability which depends on the voltage amplitude and a retention time in the range between few milliseconds to several seconds. These properties can be exploited to implement brain-inspired architectures and neuromorphic tasks, such as decision-making. In this work, we showed a possible implementation of a system able to carry out decision making tasks using volatile memristors in a multi-device synapse approach. As demonstration, we simulated the two-alternative forced choice task. The accuracy of the system was then studied under different conditions and we demonstrated the importance of tuning the switching probability and the temporal behavior of the single RRAMs to match biologically relevant timescales. Thanks to their electrically tunable characteristics in a nanoscale footprint, volatile RRAMs are very appealing candidates to be used in neuromorphic systems for decision making tasks.

## METHODS

The volatile RRAMs presented in this paper were fabricated evaporating 10 nm of hafnium oxide on top a carbon layer, and subsequent 100 nm of metallic silver as active material without breaking the vacuum. The carbon layer was the drain electrode of an integrated FET used for the 1T1R, fabricated with standard CMOS technology and used to set a compliance current ($I_{CC}$) and protect the devices from permanent short-circuits.

The quasi-static device characterization was carried out using the Semiconductor parameter analyser Agilent HP4156C. Device properties in pulsed regime were studied using an AimTTi TGA12104 Arbitrary Waveform Generator and a Tektronik MSO58 Oscilloscope. Data from the pulsed characterization, as well as set voltage distributions, retention time distributions and retention current distributions were used for the simulations. Data analysis and simulations were performed using MATLAB 2022 suit and custom codes.

ACKNOWLEDGMENT

The authors would like to thank Marco Asa, Andrea Scaccabarozzi, Claudio Somaschini, Chiara Nava, Stefano Fasoli, Stefano Bigoni, and Elisa Sogne for the help in the fabrication process. This work was partially performed in Polifab, the micro and nanofabrication facility of Politecnico di Milano. This work was supported in part by the European Union's Horizon 2020 Research and Innovation Programme under Grant Agreement No 824164 and in part by the European Research Council (ERC) through the European's Union Horizon Europe Research and Innovation Programme under Grant Agreement No 101042585. Views and opinions expressed are however those of the authors only and do not necessarily reflect those of the European Union or the European Research Council. Neither the European Union nor the granting authority can be held responsible for them.